# A Novel, Time-effective Approach for Capturing *Bacteria* from Contaminated Urine Samples

**Vincenzo Ierardi [1]\*, Paolo Domenichini [1], Domenico Vercellotti [2], and Giuseppe Vercellotti [3]**

[1] BatLab srl, Galleria Corso Garibaldi, 12/19 - 16043 – Chiavari (GE) - Italy;

[2] School of Health and Rehabilitation Sciences - Division of Health Sciences, The Ohio State University, Columbus, OH 43201;

[3] Mectron s.p.a. - Via Loreto 15° - 16042 – Carasco (GE) - Italy;

\* Correspondence: vincenzoierardi@virgilio.it

**Abstract:** A fundamental step in the race to design a rapid diagnostic test for antimicrobial resistance is the separation of bacteria from their matrix. Many recent studies have been focused on the development of systems capable of separating and capturing bacteria from different liquid environments. Herein, we introduce a new approach to this issue by using the natural bacteria tendency to accumulate at naturally-occurring interfaces, such as liquid-gas and liquid-solid interfaces, where also organic molecules like lipids, proteins, and polysaccharides accumulate. This bacterial behavior leads to the formation of a superficial layer close to the interface rich in bacteria, from which it is possible to capture a consistent amount of bacteria by means of surfaces with high chemical affinity to the outer bacteria surface. This paper demonstrates how to capture bacteria from contaminated urine samples, by means of commercial microscope slides coated with positively charged biomolecules , without recurring to bacterial culture. Moreover, this approach is an easy, quick and economical method to concentrate living bacteria in a well-defined position onto a microscope slide, thus making them easily available for further diagnostic investigations.

## 1. Introduction

A recent increase in antibiotic-resistant bacteria strains has driven researchers to develop new methods and devices that can correctly guide a physician to the most efficient antibiotic for a specific infection [1-3]. The need to rapidly identify and start treatment of infections becomes critical to avoid serious consequences for patients [4]. It is clear that the first step in creating a rapid diagnostic test for bacterial antibiotic resistance is the isolation of the bacterial strain itself. Therefore, many recent studies focused on different methods able to capture or separate bacteria from various environments, such as biological fluids, water, or food matrices[5, 6]. Indicative examples of these new approaches are the search and engineering of specific ligand that bind their targets with high affinity and specificity [7], magnetic beads [8], magnetic nanoparticles [9], microfluidic devices [10], immune-capturing techniques, and immune-magnetic separation [11]. Despite this multitude of new technologies, the global need for a simple, rapid, and inexpensive method for capturing and separating bacteria from liquid media remains. This manuscript introduces a very simple, rapid, and economical method that allows capturing and separating bacteria from urine bacterial contaminated samples. The most important difference with others methods is that our approach exploits a natural bacterial behavior. Indeed, the core feature of this method is based on the bacteria natural tendency to accumulate at the air-water interface. This phenomenon is well documented and it plays an important role in many natural environments [12, 13]. Bacteria direct their movements according to the concentration of certain chemicals in their environment (chemotaxis). Bacterial chemotaxis is what prompts the bacteria to move towards environments that contain higher concentrations of beneficial chemicals or lower concentrations of toxic substances. This is important for bacteria to find food or avoid dangerous chemicals [14]. As a matter of fact, bacteria accumulate in the superficial microlayer of liquid systems, where also food molecules such as lipids, proteins, and polysaccharides concentrate. Conceptually, it should be possible to capture bacteria from this microlayer by using a solid support functionalized with a layer of biomolecules having high affinity with the cell wall, i.e. the outer layer of the bacteria cell. For instance, it is possible to capture bacteria from this microlayer using commercial microscope slides positively charged, because the bacteria cells in physiological condition possess an overall negative charge due to the presence of peptidoglycans, which are anionic polymers [15].

Our method is based on the idea to facilitate the contact between the bacteria in the superficial microlayer and a bioactive surface putting it in rotation. To understand how the capturing method works, three aspects must be considered: the bacteria, the capturing surface, and the air-water interface. Bacteria spontaneously tend to adhere to solid surfaces by means of an initial, reversible attachment, which is followed by the transition to an irreversible adhesion to the solid surface [16]. This articulated process is facilitated by the bacteria extracellular appendages, i.e. flagella and pili, which are crucial during the initial stage of the adhesion process [17, 18]. Therefore, using a rotating system it is possible to vastly increase the number of bacteria in proximity to the capturing surface and thus trigger their adhesion to it. However, physical and chemical properties of the capturing surface affect the efficiency of bacterial adhesion [19]. Nanometric structured surfaces can greatly influence bacterial adhesion, for instance nanoscale structures or irregularities on the surface tend to increase the surface area and therefore facilitate bacterial adhesion [20]. Bacteria can spontaneously adhere to surface with a wide range of chemical properties. However, the two main factors that influence bacterial surface interactions are hydrophobicity and charge. Bacteria with hydrophobic cell surfaces prefer solid surfaces made of hydrophobic materials and vice versa. Similarly, the bacterial surface is often negatively charged thus solid surface with positive charges are more suitable for bacterial adhesion than those that have negative charges [21]. Indeed, the physical and chemical properties of the surface are fundamental to the efficiency of the bacteria capturing process. It is well known that many types of cells adhere firmly to solid substrates pretreated with polylysine, as well as to positively charged surfaces [22]. Polylysine coated and positively charged microscope slides are commercial products routinely used to immobilize cells onto glass substrates for subsequent investigation. However, it is clear that any kind of substrate with a bioactive surface, which is able to facilitate bacterial adhesion, is suitable to be used as a capturing surface. Eventually, bacteria in liquid samples, such as urine, tend to move towards the air-liquid interface following a chemotactic stimulus because all nourishment is more abundant at the interface. Using polylysine coated and/or positively charged microscope slides is possible to collect bacteria from contaminated urine samples by means of a simple device, outlined in Figure 1. This article introduces a novel, time-effective, and economical approach to capturing bacteria from bacterial contaminated urine samples by taking advantage of their natural tendency to aggregate at physical interfaces. Our objectives are to: 1) Introduce the novel experimental setup, with a discussion of the factors that affect its performance; 2) Characterize this capturing method performance in relation to well established methods for assessing bacterial concentration; 3) Determine capture bacteria availability, as a necessary condition for using them in subsequent analyses.

## 2. Materials and Methods

*Experimental Setup*

Basically, the simple idea underlying the experimental setup is to facilitate the contact between the bacteria in the superficial microlayer and a bioactive surface. This is accomplished by means of a step motor that provides a rotatory motion to the solid support bearing the bioactive surface, which is partially immersed into the bacterial contaminated urine samples. As a result, bacteria are captured onto a region of the microscope slide precisely located at the interface between air, liquid, and solid substrate

The device, outlined in Figure 1, consists of different connected parts: a step motor that generates a rotatory motion applied to the capturing surface; a holder for the solid substrate with the capturing surface; a holder for the bacterial suspension; and a simple electronic controller to set the rotatory speed. The overall system is placed in a temperature-controlled environment.

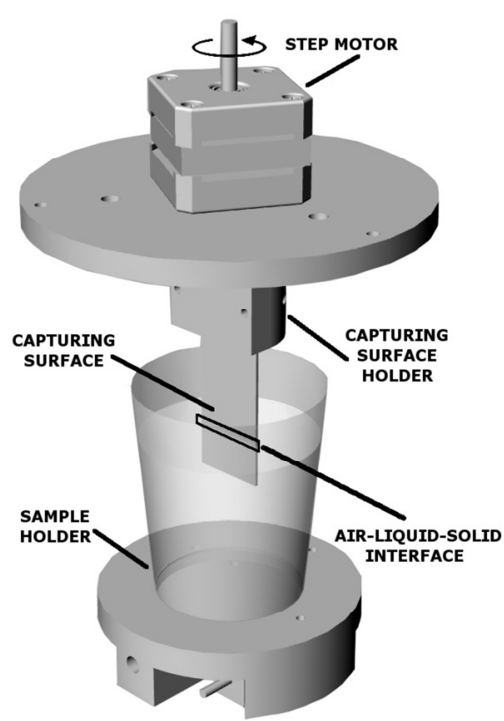

**Figure 1.** Scheme of the capturing device.

The process of capturing bacteria from liquid suspensions using the method presented here is influenced by various factors. The first and most important factor is the type of bioactive surface employed. Different types of bioactive surfaces exhibit different efficiency in promoting bacterial adhesion to them, so it is essential to use surfaces with high chemical affinity to the bacterial wall. However, even for low efficient bioactive surfaces, it is possible to optimize other factors to maximize their bacteria capturing ability. For any chosen bioactive surface, the number of bacteria captured is influenced by the rotation speed of the capturing surface, the capturing time, the temperature of the sample, and finally by the concentration of bacteria in the sample.

The experiments reported here were performed with two types of commercial microscope slides: Thermo Scientific™ Polysine (Fisher Scientific - Milan - Italy) adhesion slides that are electrostatically and biochemically adhesive; and the positively charged Klinipath (Avantor delivered by vwr – Milan

- Italy) slides. These slides are commercial products for histological applications and they have a certain grade of chemical affinity for cells. Both types of slides have shown similar capability to capture bacteria from their suspension with our system. Prior to their use, the slides were examined with an optical microscope to verify the homogeneity of their surface, i.e. an extremely smooth surface not exhibiting any type of structure visible under a 40x optical objective lens. The slides that did not pass this visual inspection were discarded. Subsequently, the slide was mounted onto a holder, which is connected to the step motor, and partially immersed in a sample of 70 mL of urine, paying attention to avoid bubble formation close to the slide surface. In the capturing phase, the following parameters were set: rotational speed of 60 rpm; capturing temperature 25 ± 1 °C, and capturing time of 30 minutes.

### *Samples and Assessment of Bacterial Concentration*

The contaminated urine samples analyzed in this experiment were obtained from subjects with urinary tract infection (UTI) caused mostly by *Escherichia coli*. The samples' bacteria concentrations were also determined with other two techniques: by means of bacterial culture tests following the guideline of the European Association of Urology [23] and direct microscopic count using a Bürker-Türk counting chamber, which has a volumetric grid divided into differently-sized cubes useful for accurately counting the number of bacteria in a cube by means of an optical microscope, and then calculating the concentration of the entire sample. In Figure 2 is displayed a capturing system working with three capturing surfaces simultaneously. After 30 minutes of rotation at 60 rpm, the slides were removed from the suspension and washed with 1 ml of MilliQ water obtained with a Millipore system, then gently dried with a flow of nitrogen, and lastly examined with an optical microscope using first a 5x optical objective lens to detect the whole bioactive surface, which presents as a transverse band across the slide. The washing step is essential in order to obtain a reliable bacterial count in the following phase of bacterial density determination. Since urine samples are rich in salts and environmental debris, washing the slides after the capture phase cleanses the bacteria band from most of the debris that may be confused with bacteria during the counting phase. The effectiveness of the washing step was validated by comparing 3D topographic images of the captured bacteria before and after the washing step. The 3D images were collected with an Atomic Force Microscope (AFM) Dimension 3100 Veeco. The AFM images analysis confirmed both the effectiveness of the washing step and that the band identified with the optical microscope consists of bacteria, whose shapes are easily recognizable in the AFM images (see Figure 5). Subsequently, a set of at least 30 optical images along the bacteria band were collected using a 40x optical objective lens and a digital camera connected to the microscope. These images were analyzed to evaluate bacterial density on the surface by counting the bacteria in the frame of the images by means of ImageJ, a freeware image processing software [24]. The surface density of bacteria obtained with the experimental setup was compared with the concentration of bacteria obtained using the bacterial culture method and the direct microscopic count method, to assess whether a relation between the amount bacteria captured and their concentration in the sample occurs.

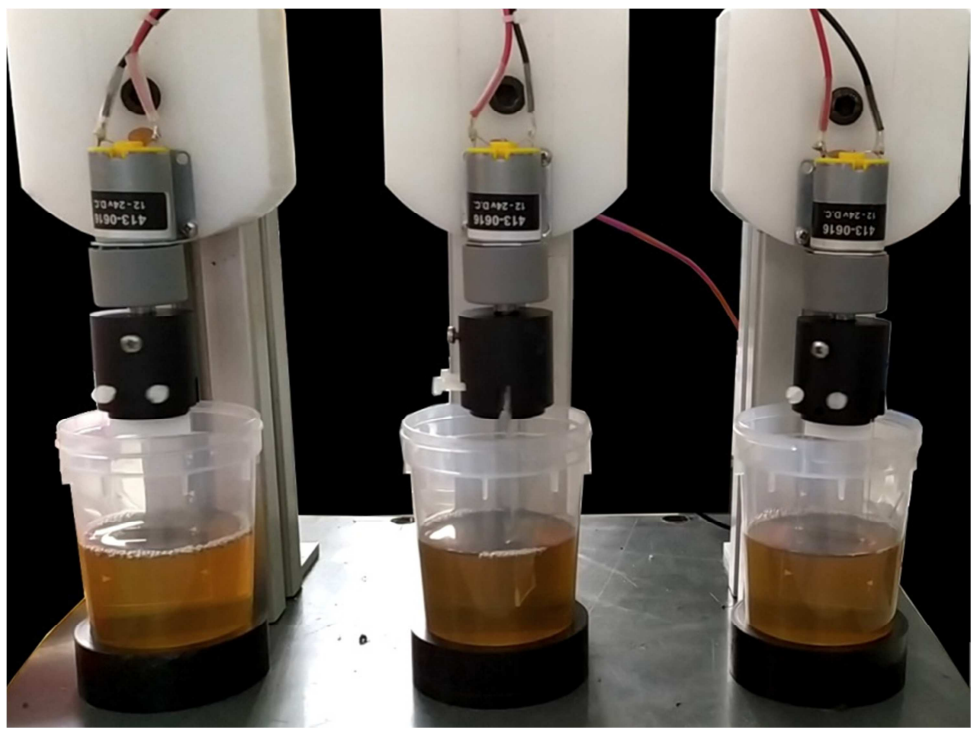

**Figure 2.** A capturing device prototype with three working lines. In the image are clearly visible the urine samples and the capturing slides connected to the motors axes through the slides holders.

*Bacterial Viability Control*

The study of direct effect of drying processes on microorganism has revealed high tolerance of bacteria to these processes [25]. To verify that the bacteria were alive after capture, a bacteria growing test was implemented. Two sets of slides were sterilized with bleach. The first series of slides were used to capture bacteria from contaminated urine samples, while the other series of slides were used in blank capturing experiments, i.e. using sterile samples. Those experiments were performed in parallel and handled in the same way and under highly controlled, sterilized conditions in order to avoid any environmental contamination. Sample sterilization was accomplished using a biological safety cabinet equipped with UV lamps. After the capturing step, the slides were placed in 50 mL sterile tubes filled with highly nutrient liquid medium (TSB) and incubated overnight at 37 °C. The presence of viable bacteria was based on the growth of the captured bacteria and on the turbidity of the culture broth.

## 3. Results

*Assessment of Bacterial Concentration in Urine Samples*

The most remarkable result emerging from the optical images analysis, collected under 5x magnification, is a clearly visible, bright band that runs across the slide and parallel to the free surface of the urine samples (Figure 3).

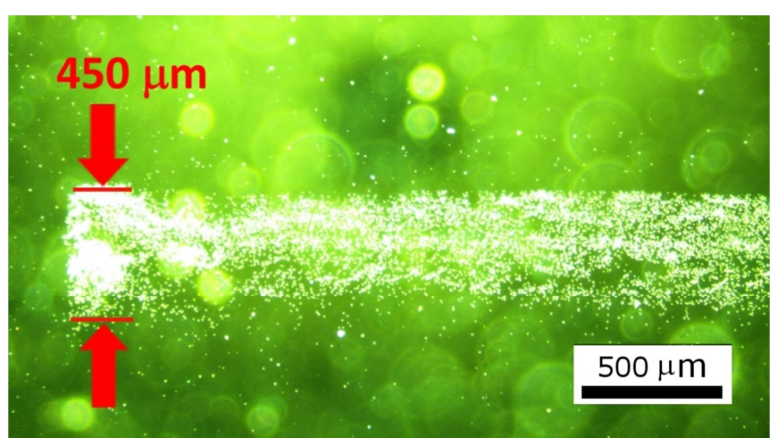

**Figure 3.** Band of bacteria obtained from a urine sample with a concentration of 15 million of bacteria per mL. Optical images collected with a 5x objective lens.

The band is made of bacteria captured from the urine contaminated samples, easily identifiable in the zoomed round areas (see Figure 4 (a), Figure 4 (b)) obtained under a 40x optical objective lens. Moreover, with a fixed set of capturing parameters, the width of the bacteria band varies in relation to the bacterial concentrations in the sample. It can range from a few tens to several hundreds of micrometers. The images collected along the band with a 40x optical objective lens were used to measure the average bacteria surface density. All images have a frame size of 250 µm x 140 µm, thus it was possible to determine bacterial density in terms of number of bacteria per frame.

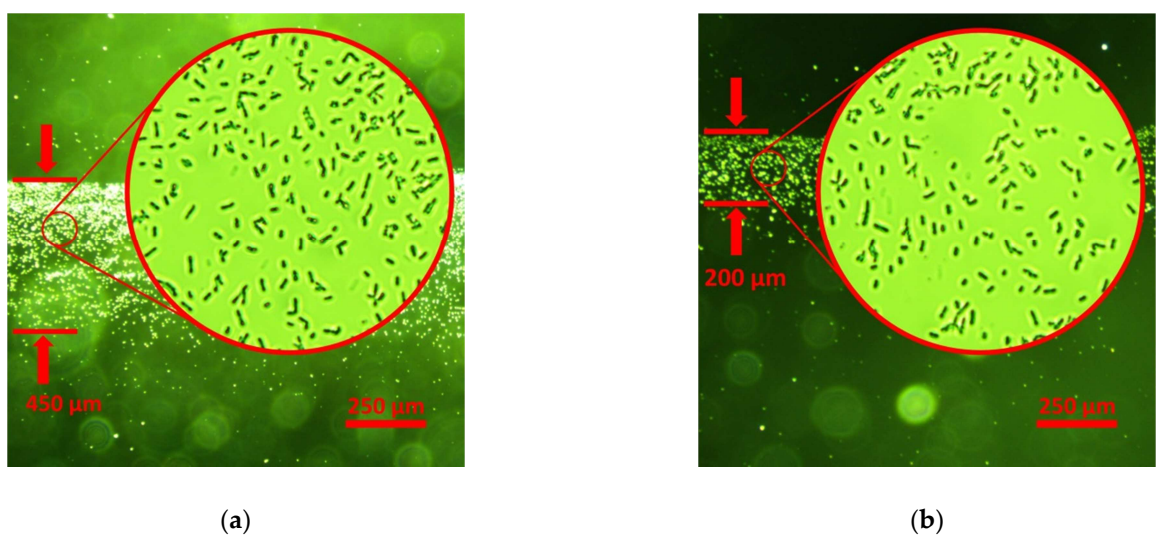

(**a**)            (**b**)

**Figure 4.** Optical images collected with a 5x objective lens (40x objective lenses for the zoomed area) in correspondence of the interface of air, liquid and capturing surface. In both images are clearly visible the bacteria attached to the surface. (a) Band of bacteria obtained from a sample with a concentration of 15 million of bacteria per mL; the zoomed area is 50 μm of radius. (b) Band of bacteria obtained from a sample with a concentration of 4 million of bacteria per mL; the zoomed area is 50 μm of radius. Both experiments were performed with slide rotation speed of 60 rpm, capturing time of 30 minutes, and capturing temperature of 25 °C.

To verify that following the washing step the captured bacteria do not include environmental debris, an atomic force microscopy (AFM) analysis was performed. This analysis compared AFM-images collected before and after the washing step. The AFM images analysis confirmed that the particles captured into the band are bacteria, which are easily recognizable in the AFM images. The results of these measurements can be seen in Figure 5. The presence of environmental debris is evident in the AFM images collected before the washing step, highlighted by red arrows in Figure 5(a), while there is no evidence of significant presence of environmental debris in Figure 5(b) collected after the washing step. These results assure that the majority of the particles forming the band are indeed bacteria captured from contaminated urine samples. The image processing software ImageJ was used to count the number of bacteria per frame. Since bacteria captured onto the bioactive surface tend to form clusters, clusters comprising a maximum of six bacteria were also included in the counting procedure. ImageJ can count bacteria through the command "COUNT PARTICLES" on 32 bit black and white images and it allows setting the area range of the particles. This procedure is similar to the one reported in a previous work [26].

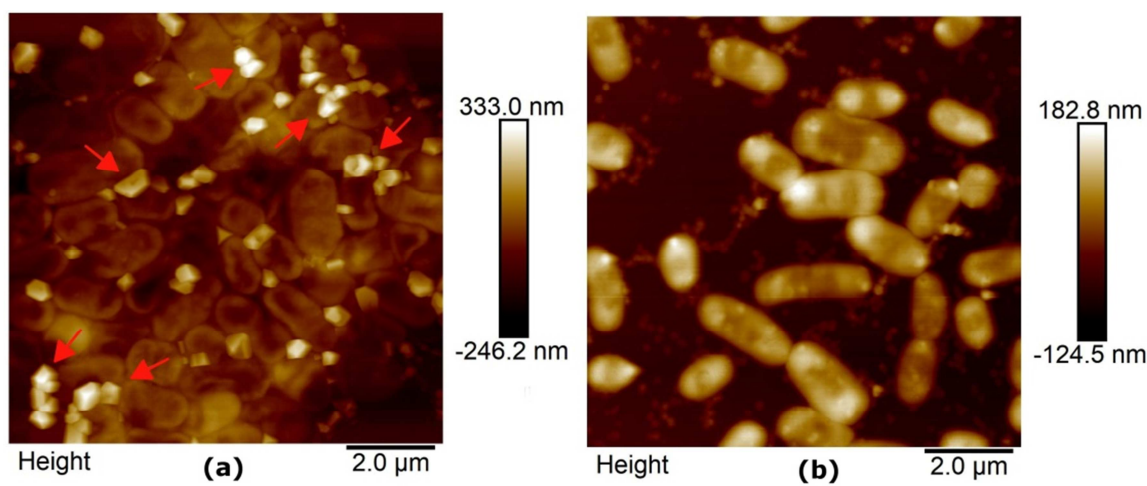

**Figure 5.** AFM images of the captured bacteria: (a) before the whasing step, the arrows indicate enviromentel debris, likely salt cristals and (b) after the washing step, there is no evidence of significative debris.

The average of the bacteria surface density (bacteria/frame), i.e. the average number of bacteria captured per image, is used to verify if a correlation between the bacteria concentration in the urine samples and the bacteria captured onto the slide surface exists. Therefore, to evaluate the existence of a possible correlation between the numbers of bacteria captured onto the slides and their concentration in urine contaminated samples, a set of 24 capturing experiments were performed. In half of the urine contaminated samples, bacterial concentration was determined by direct microscopic count method, while for the other half it was determined by bacterial culture method. For each sample, the capturing experiment was replicated twice, once using the Thermo Scientific™ Polysine adhesion slides, and once using the positively charged Klinipath slides, and the average value of the bacteria surface density was taken, since there were no significant differences in the number of bacteria captured with the two types of slides. Graphs in Figure **6** show the dependence between the concentration of the bacteria in contaminated urine samples and the bacteria surface density determined with our method. Indeed, the implementation of a calibration curve based on bacteria concentration vs bacteria surface density, allows determining the unknown bacteria concentration of contaminated urine samples. In particular, Figure 6 (a) and (b) show the relations between the bacterial concentrations obtained with the culture bacteria method and with the direct microscopic count method, and the average bacteria surface density obtained with the capturing system/method here described. Ordinary least square regression was used to determine the best fit line for both sets of data, with R-squared of 0.99 and 0.90 respectively (dashed line). Moreover, Table 1 reports the data of the two sets of measurements performed with contaminated urine samples that were used in the regression analysis. In contrast, the microscopic analysis of capturing slides from uncontaminated samples do not shown presence of bacteria. The uncertainties associated with these experiments are slightly dif-ferent: in the case of the bacterial culture method the error is around the 20% [27], while the error associated with the direct microscope count method, using a Bürker-Türk counting chamber, is around 15% [28]. The error estimated for bacteria surface density measurements is 5%, calculated from the standard error of the mean of the bacteria count data.

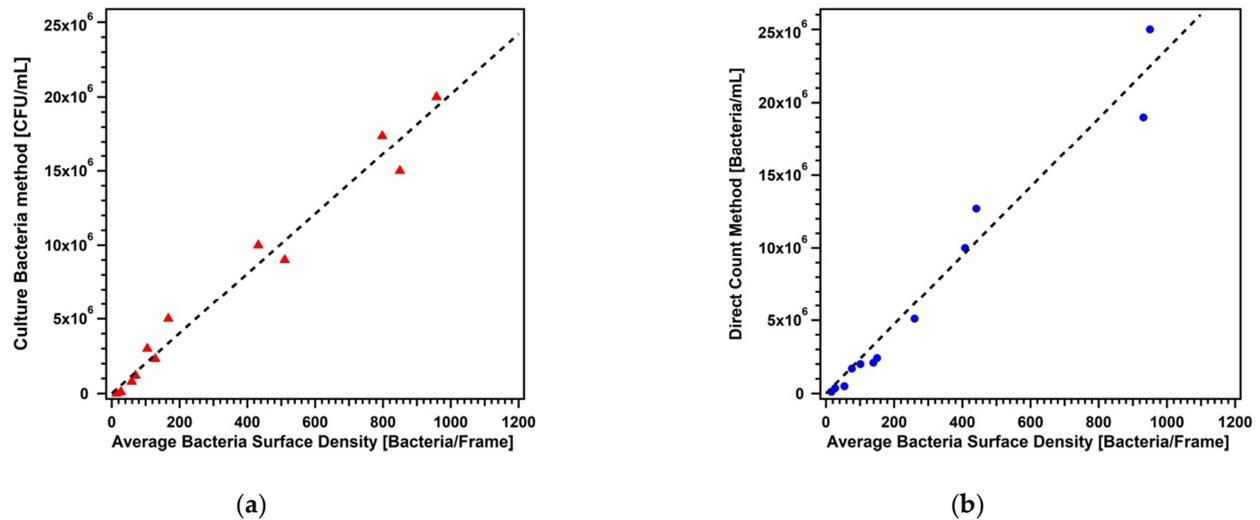

(**a**) (**b**)

**Figure 6.** Graphs of the correlation between bacteria concentrations in urine samples and the correspondent average bacteria surface density obtained from optical images collected with a 40x objective lens, slide rotation speed of 60 rpm, capturing temperature of 25 °C, and capturing time of 30 min. (a) Correlation between the concentration of bacteria in urine samples obtained by means of the culture method and the bacteria surface density obtained with the capturing method here described. (b) Correlation between the concentration of bacteria in urine samples obtained by means of direct microscopic count and the bacteria surface density obtained with the capturing method here described. Dashed lines are obtained by a linear regression of the data, with R-squared of 0.99 and 0.90 respectively.

**Table 1.** Dataset relative to contaminated urine samples.

| Bactrial Culture (CFU/mL) | Slides Capture (Bacteria/frame) | Direct Count (Bactria/mL) | Slides Capture (Bactria/frame) |
|---|---|---|---|
| $3.00 \cdot 10^{+3}$ | 14 | $1.20 \cdot 10^{+5}$ | 16 |
| $1.00 \cdot 10^{+5}$ | 27 | $3.60 \cdot 10^{+5}$ | 27 |
| $8.00 \cdot 10^{+5}$ | 59 | $5.00 \cdot 10^{+5}$ | 54 |
| $1.20 \cdot 10^{+6}$ | 70 | $1.70 \cdot 10^{+6}$ | 76 |
| $2.32 \cdot 10^{+6}$ | 128 | $2.00 \cdot 10^{+6}$ | 101 |
| $3.00 \cdot 10^{+6}$ | 105 | $2.10 \cdot 10^{+6}$ | 139 |
| $5.00 \cdot 10^{+6}$ | 167 | $2.41 \cdot 10^{+6}$ | 150 |
| $9.00 \cdot 10^{+6}$ | 510 | $5.10 \cdot 10^{+6}$ | 260 |
| $1.00 \cdot 10^{+7}$ | 432 | $1.00 \cdot 10^{+7}$ | 408 |
| $1.50 \cdot 10^{+7}$ | 850 | $1.27 \cdot 10^{+7}$ | 441 |
| $1.74 \cdot 10^{+7}$ | 798 | $1.90 \cdot 10^{+7}$ | 931 |
| $2.00 \cdot 10^{+7}$ | 958 | $2.50 \cdot 10^{+7}$ | 950 |

To properly determine the most suitable values of the capturing parameters for our apparatus, i.e. speed, time and temperature, sets of experiments were performed to examine the effects of these parameters on the efficacy of the capturing phase.

The first parameter analyzed was rotation speed. From the capture experiments where only the capturing slide rotational speed was changed, it is evident that a directly proportional effect exists between speed and capture efficacy, below 150 rpm. Indeed, speeds higher than 150 rpm produce irregular bands of captured bacteria, as shown in the optical images of different bands of captured bacteria reported in Figure 7. Rotational speeds comprised between 30 rpm and 160 rpm lead to an evident increase of the superficial density of captured bacteria, as illustrated in Figure 7D to Figure 7B; while at higher rotational speed are evident irregularities of the bacterial surface density, see Figure 7A. Therefore, a rotational speed between 50 rpm and 70 rpm seems to be the most suitable for our apparatus, whose target speed was set to 60 rpm. This rotational speed exhibited a positive correlation between the suspension bacterial concentration and the surface density of bacteria captured.

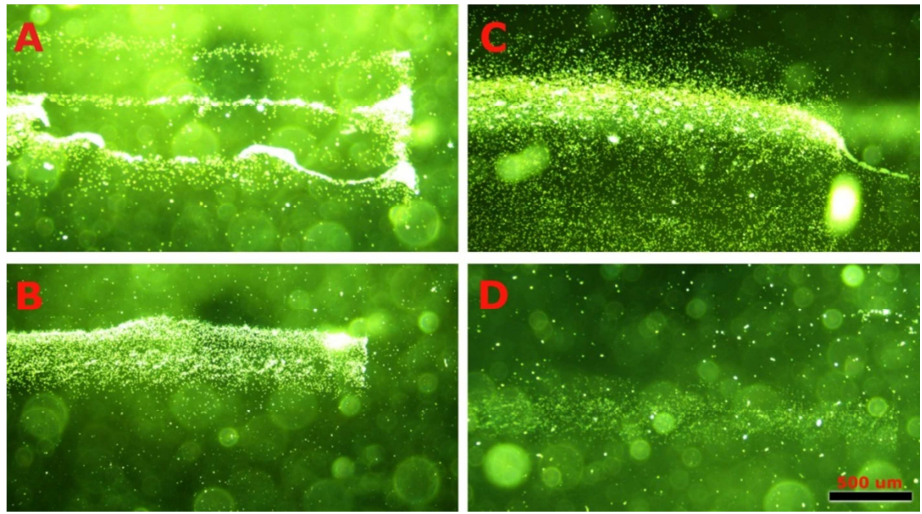

**Figure 7.** Example of the bacteria captured bands obtained at four different capturing speeds from a urine sample with a bacteria concentration of 8.3 million CFU/mL at room temperature and 60 min. of capturing time. Optical images collected with a 5x objective lens: A) capturing speed of 160 rpm; B) capturing speed of 70 rpm; C) capturing speed of 50 rpm, and D) capturing speed of 30 rpm.

Additionally, it was observed that capturing time influences the number of bacteria captured, with greater numbers of bacteria being captured by the bioactive surface over longer capturing times. Nevertheless, a short capturing time is desirable to devise a quick, reliable method, thus making it necessary to find a balance point between the amount of bacteria captured and the time necessary to capture them. To find the most suitable capturing time, three bacterial suspensions with concentrations of $2.0 \cdot 10^5$ CFU/mL, $1.5 \cdot 10^6$ CFU/mL, and $8.2 \cdot 10^6$ CFU/mL, were used in a set of capturing experiments. Capturing temperature and rotational speed were held constant in all experiments, while capturing time was progressively reduced from 120 min., to 60 min., to 30 min., and to 15 min. The results of these tests are displayed in Figure 8. In all cases where capturing time was greater than 15 min. it was possible to detect a clear band of captured bacteria, whereas the 15 min. capturing time did not allow for the formation of a captured bacteria band in two less concentrated suspensions. Figure 8 shows that the increase of capturing time produces a consequential increase in the amount of captured bacteria, and the time suitable for a rapid capturing phase can be chosen in the range of 30 – 60 minutes. Moreover, from tests performed at a longer capturing time, it clearly emerges that a sort of saturation point of the bioactive capturing surface is reached, that is, after 200 min. there are no further increases in bacteria surface density.

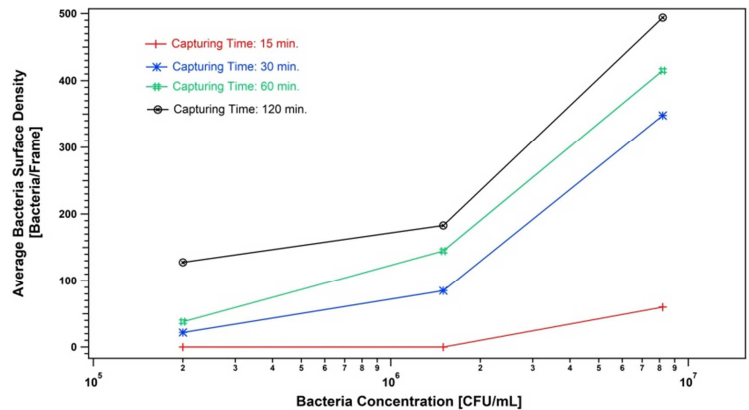

**Figure 8.** The Graph displays the variation of the bacteria surface density against the suspensions concentration relatively to four different capturing times.

Finally, the capturing temperature influences positively the amount of bacteria captured [29, 30], i.e. increasing the temperature from 20 °C to 40 °C produces an increase on the amount of captured bacteria as it is possible to see in Figure 9.

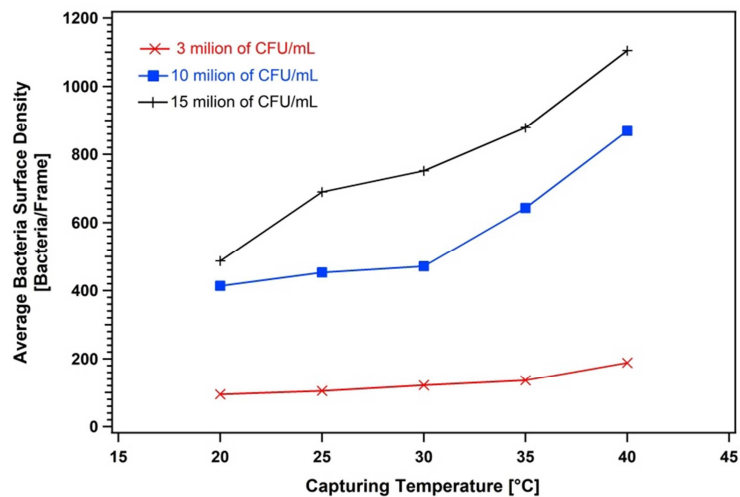

**Figure 9.** Graph of the variation of the surface density of captured bacteria, for three different bacterial concentrations, relatively at the increase of the temperature from 20 °C to 40 °C.

Likely this increase is due to two different effect: the first is that the increase in temperature produces an increase in convective thermal motion inside the bacterial suspension; and the second is associated with the bacterial growth which is higher at 40 °C than at 20 °C. Therefore, there is an increase of bacteria number in the suspension at 40 °C during the capturing phase. The capturing temperature of 25 ± 1 °C, close to the room temperature, has the double advantage of limiting the environmental thermal fluctuations and the bacterial growth.

*Bacterial Viability Control*

Following the procedure previously described, the set of slides used to capture bacteria from contaminated samples showed bacteria growth, while the set used in the blank experiments did not show any trace of bacterial growth (Figure 10). The sterilized environmental conditions in which all experiments were performed exclude that the cloudy appearance is due to external bacterial contamination. Indeed contamination would have led to bacterial growth also in the blank experiment samples, since the two sets of experiments were performed in parallel. Therefore, this result confirms bacterial viability after capture. Moreover, samples from both experiments were analyzed by means of optical microscope and the presence of bacterial growth was confirmed only for the slides used with the contaminated samples, while no presence of bacteria growth was detected in the blank experiments.

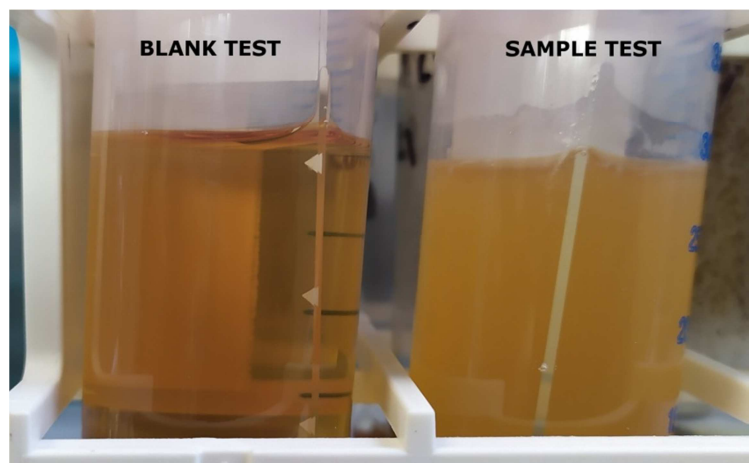

**Figure 10.** Bacteria viability verified by a growing test. In the left tube the culture broth is clearly transparent (blank), while in the right tube the cloudy culture broth is indicative of bacterial growth.

## 4. Discussion and Conclusions

Our method, at the moment, has been applied only to contaminated urine samples and the captured bacteria were classified as typical microbes that cause urinary tract infections (UTI), which are mostly *Escherichia coli* but we cannot exclude the presence of others type of bacteria responsible for UTI. However, the capability to localize precisely the bacteria captured is a strategic, extremely useful advantage that subsequently enables the utilization of characterization techniques directly on the captured bacteria, for instance Atomic Force Microscopy, Raman spectrometry, or genomic approach methods, avoiding the time consuming bacterial culture step [26, 31, 32]. The position of the captured bacteria band onto the slide is determined by the position of the slide in relation to the liquid free surface. Therefore, the presence of the bacteria onto the capturing slide is very easy to detect and to verify.

The bacteria band captured (see Figure 3) become less visible at low bacteria concentrations, even though it may still be possible to spot it. For example, a band obtained from a sample with 3000 (Colony-Forming Unity)/mL of bacteria produces an average bacteria density of 14 Bacteria/Frame. Even though in uncontaminated sample of urine the bacteria band is invisible, it is still possible to recognize a thin straight line in correspondence of the free surface of the urine sample made of sediments even if they are smaller than the bacteria, such as protein and crystals, which are normally present in urine samples. The width of the captured bacteria band basically depends on the bacteria concentration in the sample, higher concentrated sample produces a wider band compared to a sample with lower concentration, as in Figure 4(a) and (b). It is interesting to observe that, in each experiment, the average bacteria surface density decreases quickly along the vertical axis of the slide, going down from the band zone towards the bottom of the slide. For instance, in the case of a sample of urine with a concentration of 10 million of bacteria per mL, the average density of bacteria decreased from 396 (bacteria/frame) within the band to 47 (bacteria/frame) 500 micrometers below the band zone. This is consistent with the accumulation of bacteria near the free surface of the sample [33].

A well-shaped bacteria band could be a useful condition in implementing an automatic process of analysis, for instance if the slide is moved in a spectrometer for further characterization [32]. The factors that influence the characteristics of the bacteria band are mainly three: the rotation speed of the slide, the capturing time, and the capturing temperature.

The slide rotation speed influences the regularity of the band, i.e. high capturing speeds create turbulence within the sample's free surface that produce an undulated band. Thus, by setting the rotation speed in order to have the free surface of the sample as steady as possible, the bacteria are captured in a straight line. Therefore, it is needed to tune the rotation speed in order to have a well-shaped band with the bacteria surface density as high as possible. For standard cylindrical urine containers (120 mL), a rotation speed of 60 rpm produces well-shaped captured bacteria bands with a bacteria surface density sufficient also in the case of low concentration samples ($\leq 10^5$ CFU/mL).

The capturing time is directly linked to the number of bacteria captured during the rotation of the slide into the sample, i.e., increasing the capturing time increases consequently the bacteria surface density. The optimal capturing time needs to be adjusted by taking in account all experimental conditions.

Lastly, the effect of an increase of the capturing temperature, compatible with the bacteria life, induces an increase of the bacteria activity and motility [30], thus much more bacteria are able to reach the surface layer rich in nourishments producing an increase of the numbers of captured bacteria. Consequently, a decrease of the temperature has an opposite effect.

In conclusion, analyzing the data obtained from the capturing tests it is clear that there is a significant positive correlation between the concentrations of bacteria in the samples of contaminated urine and the relative bacteria surface density obtained with our method. This correlation is obtained because the capturing parameters have been fixed at the same value for the whole set of measurements, i.e. the capturing speed to 60 rpm, the capturing time to 30 minutes, the capturing temperature to 25 °C, with the same bioactive surfaces. Moreover, the results of the bacteria viability tests further strengthened our conviction that could be possible to use bioactive surfaces to capture bacteria for conducting specific antimicrobial tests directly on the capturing slides without the time-consuming step of bacterial culture.

**Author Contributions:** Conceptualization, V.I. and P.D.; formal analysis, V.I. and P.D.; investigation, V.I. and P.D.; resources, D.V. and G.V.; writing—original draft preparation V.I.; writing—review and editing, G.V.; supervision, D.V. and G.V.; project administration, D.V. and G.V. All authors have read and agreed to the published version of the manuscript.

**Funding:** This research was funded by Mectron S.p.A. , Carasco (GE), Italy.

**Acknowledgments:** The authors thank the staff of the Studio Associato Chemilab (Chiavari - Italy) for their excellent technical support.

**Conflicts of Interest:** The funder has no role in the active design of the study or in technical choices made throughout the entire project, in the collection, analyses, or interpretation of data, in the writing of the manuscript, or in the decision to publish the results.